\documentstyle[psfig,amssymb]{europhys}

\def\And{{\rm and\ }}

\def\stars{\bigskip\centerline{***}\medskip}

\newif\ifboo \boofalse

\def\mif{{\hspace{0.5cm} \rm if} \hspace{0.5cm}}
\def\mwith{{\hspace{0.5cm} \rm with} \hspace{0.5cm}}

\def\mpunkt{{\hspace{0.5cm} \rm .}}
\def\mkomma{{\hspace{0.5cm} \rm ,} \hspace{0.5cm}}

\begin{document}
\euro{26}{2}{241}{1994}
\Date{??}
\shorttitle{W. Ebeling and T. P\"oschel, Entropy and Long range correlations in literary English}

\title{Entropy and Long range correlations in literary English}

\author{Werner Ebeling\inst{1} \And Thorsten P\"oschel\inst{1,2}}
                                                                               
\institute{ 
\inst{1}Institut f\"ur Theoretische Physik, Humboldt--Universit\"at
  zu Berlin, Invalidenstra\ss e 110, D-10099 Berlin\\
\inst{2}HLRZ, Research Center J\"ulich, D-52425
J\"ulich, Germany
}


\pacs{
\Pacs{}{}{} 
}

\maketitle
\begin{abstract}
  We investigated long range correlations in two literary texts, Moby
  Dick by H.~Melville and Grimm's tales. The analysis is based on the
  calculation of entropy like quantities as the mutual information for
  pairs of letters and the entropy, the mean uncertainty, per letter.
  We further estimate the number of different subwords of a given
  length $n$. Filtering out the contributions due to the effects of
  the finite length of the texts, we find correlations ranging to a
  few hundred letters. Scaling laws for the mutual information (decay
  with a power law), for the entropy per letter (decay with the
  inverse square root of $n$) and for the word numbers (stretched
  exponential growth with $n$ and with a power law of the text length)
  were found.
\end{abstract}

From a formal point of view a book may be considered as a linear
string of letters. In this respect there exists a similarity to other
linear structures~\cite{ebelingengelfeistel}. Usually the strings
generated by dynamical systems show only short range correlations,
except under critical conditions where, in analogy to equilibrium
phase transitions \cite{stanley}, correlations on all scales may be
observed~\cite{ebelingnicolis92}.  Recently several studies on long
range correlations in DNA sequences~\cite{peng} and in human
writings~\cite{schenkel92} have been published. The intrinsic
difficulties connected with the analysis of long range correlations in
DNA led to a controversial discussion about the authentic character of
long range structures in DNA~\cite{likaneko}.
\par
This work is devoted to the investigation of long range correlations
in texts. We use the methods of entropy analysis, which were first
applied to texts by Claude Shannon in 1951 \cite{shannon}. For several
reasons we expect the existence of long range structures in these
sequences. Since a book is written in a unique style and according to
a general plan of the author, we expect correlations which are ranging
from the beginning of a text up to the end~\cite{epa}.
\par
Another strong argument for long correlations is based on the
combinatorial explosion. Uncorrelated sequences generated on an
alphabet of $\lambda$ letters have a manifold of $\lambda^n$ different
subwords of length $n$. A subword (block) is here any combination of
letters including the space, punctuation marks and numbers. For $n
>100$ the number $N(n)$ is extremely large.  Hence we we must expect
that only a very small subset $N^*(n)$ of the possible words appears
in a text. Bounds of this kind are given by the rules of writing
texts, i.e. by the rules of syntax as well as by semantic relations,
which do not allow for an arbitrary concatenation of letters to words
and of words to sentences. The problem we address here is, whether the
function $N^*(n)$ follows a simple scaling law.
\par
In earlier papers the conjecture has been made that the number of allowed
subwords scales according to a stretched exponential 
law~\cite{ebelingnicolis92,hilberg90}
\begin{equation}
  N^{*}(n)\sim \exp[cn^{\alpha}] \mwith \alpha <1 \mkomma c=const.
\label{scalingrule}
\end{equation}
The the scaling rule (\ref{scalingrule}) reduces the number of the
allowed subwords drastically ($N^*(n) \ll \lambda^n$) for large $n$.
In order to describe a given string of length $L$ using an alphabet of
$\lambda$ letters we introduce the following
notations~\cite{ebelingnicolis92}: Let $A_1 A_2 \dots A_n \nonumber$
be the letters of a given substring of length $n\le L$. Let further
$p^{(n)}(A_1\dots A_n)$ be the probability for this substring (block).
A special case is the probability to find a pair with $(n-2)$
arbitrary letters in between $p^{(n)}(A_1,A_n)$.  Then we may
introduce the mutual information for two letters in distance $n$
\cite{herzel,li}:
\begin{equation}
  I(n)=\sum_{A_i A_j} p^{(n)}(A_i,A_j)\log \left[
  \frac{p^{(n)}(A_i,A_j)}{p^{(1)}(A_i)\cdot p^{(1)}(A_j)}\right]
\end{equation}
which is closely related to the autocorrelation
function~\cite{peng,herzel,li,nicoliskatsikas}. Further we define the
entropy per block of length $n$~\cite{grassberger90}:
\begin{equation}
  H_n=-\sum p^{(n)}(A_1 \dots A_n) \log p^{(n)}(A_1 \dots A_n) \mpunkt
\end{equation}
The block entropy is related to the mean number of
words~\cite{ebelingnicolis92} by
\begin{equation}
  N^{*}(n)\sim \lambda^{H_n} \mpunkt
\label{mcmillanlaw}
\end{equation}
As shown already by Shannon $H_n/n$ is an important quantity
expressing the structure of sequences.  In~\cite{ebelingnicolis92} we
assumed the scaling
\begin{equation}
\begin{array}{l}
  H_n/n=h + g\cdot n^{\mu_0-1} + e/n \\ 0\le \mu_0 <1 \mkomma
  n\rightarrow\infty \mpunkt
\end{array}
\label{ebelingscaling}
\end{equation}
Here $h$, the limit of the mean uncertainty, is called the entropy of
the source. This quantity is positive for stochastic as well as for
chaotic processes, $g$ and $e$ are constants; if $h,e>0$ and $g=0$ the
correlations in the string are short range corresponding to a Markov
process with a finite memory~\cite{grassberger90}. For periodic
strings one finds $h=g=0$, $e>0$.  The existence of a long range order
in strings may be characterized by the condition $g>0$ describing a
slowly decaying contribution to the asymptotics of the entropy per
letter for large $n$. Of special interest for the further
consideration of texts is the case $h \ll 1$, $g>0$ corresponding to a
power law tail of the entropy decaying slower than $1/n$. It would be
interesting and important to estimate the limit entropy $h$ for
``homogeneous texts'', however, there are not enough data to do it
with sufficient reliability. In the following we assume that
$h\approx0.01\dots 0.1$.  Therefore it may be neglected in our
investigations which are restricted to $n<30$.
\par
The mutual information is not a monotonic function of $n$. We define
long range effects by power law tails of the averaged mutual
information $I(n)$. Here the averaging is carried out over a window
comprising several of the typical oscillations (fluctuations). Several
authors have demonstrated that DNA-sequences show a slowly decaying
fluctuations at large scales
~\cite{herzel,li,nicoliskatsikas,herzelschmittebelingcsf,voss}.
\par
We will apply the methods of entropy analysis to literary English
represented by the books: ``Moby Dick'' by Melville ($L\approx
1,170,200$) and Grimm's Tales ($L\approx 1,435,800$).  Pieces of music
may be treated in a similar way~\cite{epa}.  For simplification we use
an alphabet consisting of $\lambda=32$ symbols: the small letters {\it
  a $\dots$ z} \hspace{0.2cm} the marks {\it , . ( ) \# } and the
space; {\it \#} stands for any number. In order to get a better
statistics we have used for the entropy calculations also a restricted
alphabet consisting of only $\lambda=3$ letters {\it 0, M, L}.  {\it
  0} codes for vowels, {\it M} for consonants and {\it L} for spaces
and marks.
\par
\begin{figure}[ht]
  \centerline{\psfig{figure=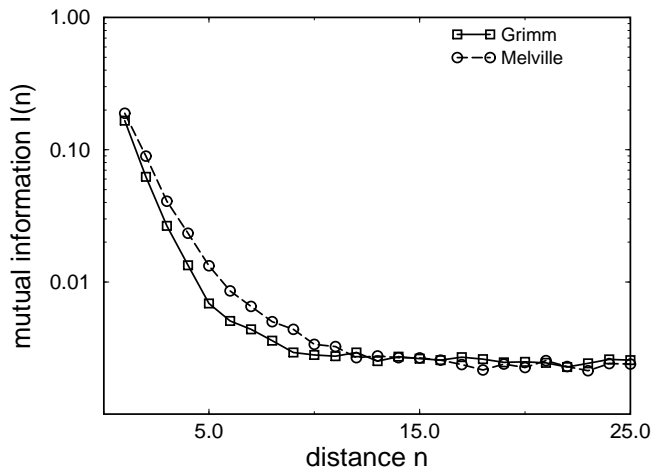,width=8cm,angle=270}}
  \caption{\it The mutual information calculated for  Melville's Moby Dick 
    and for Grimm's Tales ($\lambda=32$).}
\label{mutualmobygrimm}
\end{figure}

To estimate the mutual information we count frequencies of pairs of
letters at distance $n$.  Fig.~\ref{mutualmobygrimm} shows the mutual
information calculated for Moby Dick and for Grimm's Tales
($\lambda=32$). The results show well expressed correlations in the
range $n=1\dots 25$ which are followed by a long slowly decaying tail.
The obtained values for the transinformation $I(k)$ become meaningless
if they are smaller than the level of the fluctuations $\delta I(k)$
due to the finite length $L$ of the
text~\cite{herzel,herzelschmittebelingcsf}:
\begin{equation}
  \delta I(k)=\frac{\lambda^2-2\cdot\lambda}{2\cdot \ln\lambda\cdot
    L} \mpunkt
\end{equation}
For our rather long texts with $L>10^6$ the fluctuation level has a
value of about $10^{-4}$. The smoothed values for the mutual
information for the range $n=25\dots 1000$ may be fitted by the
scaling law $I(k)=c_1\cdot n^{-0.37}+c_2$ with $c_1=1.5\cdot 10^{-4}$,
$c_2=1.1\cdot 10^{-4}$. The constant $c_2$ corresponds here to the
level of fluctuations.  Our results show that long texts show pair
correlations which decay, at least up to distances of several hundred
letters, according to a power law, however, in the range $n=100\dots
1000$ the fluctuations are rather strong and the mean square deviation
reaches $20\dots 40\%$.
\par
\begin{figure}[ht]
  \centerline{\psfig{figure=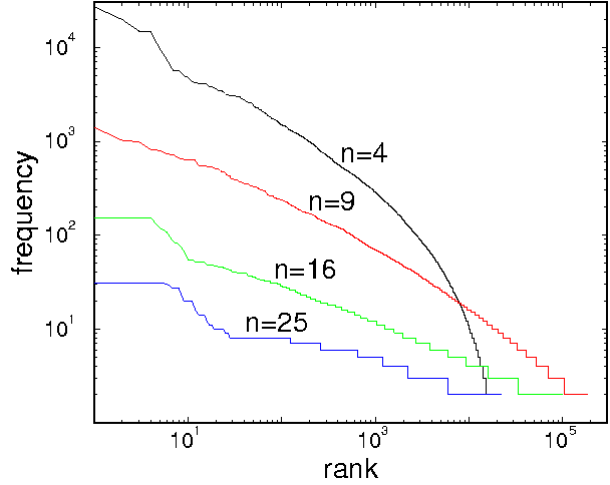,width=8cm}}
  \caption{\it The observed rank ordered distribution of words of length 
    $n=4, 9, 16, 25$ for Grimm's Tales.}
\label{rankgrimm}
\end{figure}

For the calculation of entropies we count the frequencies of subwords,
where a subword of length $n$ is defined as any combination of $n$
letters. The results for $n=4,~9,~16,~25$ letters in Grimm's Tales are
shown in Fig.~\ref{rankgrimm} in a rank ordered representation.  The
structure of the rank ordered distributions is for both texts rather
similar, however the list of words is of course very different.
\par
The form of the subword distributions is distinctly not Zipf--like, it
does not follow a power law. In the opposite, with increasing $n$
there is a tendency to form a plateau~\cite{epa}. The effects due to
finite $n$ and the effects of finite length $L$ tend to smooth the
edges of the distribution~\cite{herzelschmittebelingcsf}.  The
importance of length corrections for estimating the frequencies of
subwords was considered by several
authors~\cite{herzel,grassberger90}. For a deeper analysis of this
problem we refer to recent
articles~\cite{herzelschmittebelingcsf,schmittherzelebelingepl,pengbuldyrev}.
Our method for the entropy analysis uses an extrapolation of the
entropy to infinite text length~\cite{ebelingnicolis92}.
\par
The probabilities which we need for the calculation of entropies are
unknown and can only be estimated from the frequencies $N_i(n)$ of the
subwords of length $n$ in a text of length $L$ containing $N=L+1-n$
subwords.  Introducing the observed subword frequencies into the
entropy definition leads to the observed entropies
\begin{equation}
  H^{obs}_n=\log(N)-\frac{1}{N}\sum_iN_i(n)\cdot
  \log\left(N_i(n)\right) \mpunkt
\end{equation}
This is a random variable with the expectation value
\begin{equation}
  H^{exp}_n=\langle H^{obs}_n\rangle=\log(N)-\frac{1}{N}\sum_i\langle
  N_i(n)\cdot \log \left(N_i(n)\right)\rangle \mpunkt
\end{equation}
Assuming a Bernoulli distribution for the letter combinations, the
mean values can be calculated
explicitly~\cite{herzel,herzelschmittebelingcsf}. The result is
\begin{equation}
  H^{exp}_n= \left \{
  \begin{array}{ll}
    H_n-\frac{N^*(n)}{2 N} & \mif N^*(n) \ll N \\ \log (N)-\log (2)
    \frac{N}{N^*(n)} & \mif N^*(n) \gg N \mpunkt
  \end{array} 
\right.
\label{approximation}
\end{equation}
The latter case corresponds to the situation that only a minor part of
the possible subwords of length $n$ have a chance to appear in the
text.
\par 
The relation between the effective number of words $N^*(n)$ and the
block entropy $H_n$ is given by eq.~(\ref{mcmillanlaw}).  Hence the
expected block entropy may be represented as a function of $\log N$
with one free parameter $H_n$ which is found by fitting the curves.
In this way the block entropies for both books were calculated up to
$n=26$.  For small word length i.e. $n \lesssim 16$ for $\lambda = 3$
and $n\lesssim 5$ for $\lambda = 32$, we used the approximation
(\ref{approximation}) for $N^*(n)\ll N$. For larger $n$, i.e.
$n\gtrsim 20$ for $\lambda = 3$ and $n\gtrsim 10$ for $\lambda = 32$,
we applied the approximation (\ref{approximation}) valid for
$N^*(n)\gg N$.  More concrete, we measured the deviation between the
observed entropy and $\log(N)$. Then the theoretical entropy $H_n$ was
estimated from $N^*(n)$ using eq.~(\ref{mcmillanlaw}).  In the
intermediate region we applied a smooth Pad\'{e} approximation between
both formulae.  In a procedure of successive approximations the
entropy $H_n$ was considered as a free parameter which was fitted in a
way that $H^{exp}_n (\log N)$ came as close as possible to the
measured (observed) entropy values. In practice this method breaks
down for $n\gtrsim 30$ if $\lambda = 3$ and for $n\gtrsim25$ if
$\lambda = 32$. Longer subwords do not have a chance to appear several
times in the text, what leads to large statistical errors.
\par
\begin{figure}[ht]
  \centerline{\psfig{figure=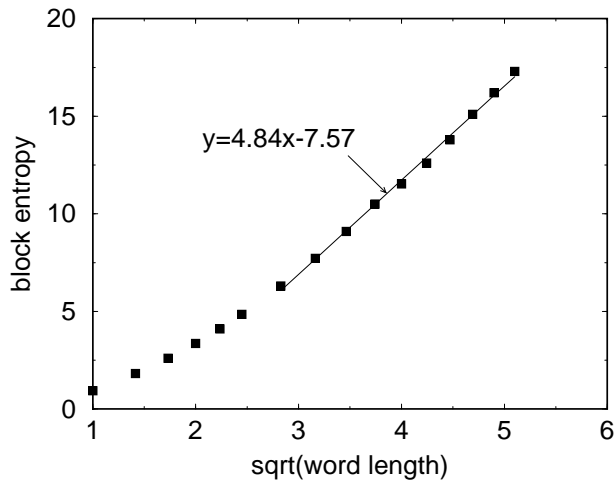,width=8cm,angle=270}}
  \caption{The scaling behavior of the block entropy $H_n$ with the square root of
    the word length $n$ for Moby Dick encoded by the alphabet
    $\lambda=3$.}
\label{mobyentr}
\end{figure}

The calculations for $n \le 26$ show that the square root law yields a
reasonable approximation for the scaling of the entropy per letter
with the word length $n$
\begin{equation}
\begin{array}{llllll}
  H_n/(n \cdot \log(\lambda)) & \approx & (4.84/ \sqrt{n}) & -(7.57/n)
  & \hspace{0.1cm}(\lambda =3) \\ 
  H_n/(n \cdot \log(\lambda)) &
  \approx & (0.9/ \sqrt{n}) & + (1.7/n) & \hspace{0.1cm}(\lambda
  =32)\mpunkt
\end{array}
\end{equation}
Fig.~\ref{mobyentr} shows the fit for the alphabet $\lambda=3$.  The
scaling law of the square root type was first found by
Hilberg~\cite{hilberg90} by fitting Shannons original data. For
$n=100$ and $\lambda=32$ our scaling formula yields $H_{100}\approx
10\cdot \log(\lambda)$ what is not far from Shannon's estimation
$H_{100}\approx 40$ bits.
\par
The number of subwords increases according to a stressed exponential
law. For the growth we found the approximation
\begin{eqnarray}
  N_n^*\approx & 2^{7.45 \sqrt{n}-11.3} & \hspace{1cm}(\lambda=3)\\  
  N_n^*\approx & 2^{4.5 \sqrt{n}+8.5} & \hspace{1cm} (\lambda=32)
  \mpunkt
\end{eqnarray}
We summarize now the results obtained for the two books: The scaling
of the mutual information and the entropy per letter shows in
agreement with earlier work~\cite{ebelingnicolis92} that long texts
are neither periodic nor chaotic but somehow in between. Taking into
account length corrections we calculated block entropies up to $n=26$
and mutual information values up to distances of a few hundred
letters. Based on these data we formulated a hypothesis about the long
range scaling.  For the range $n \gtrsim 100$ the pair correlations
contained in the transinformation of long texts $L>10^6$ decay
according to a power law, however the level of fluctuations is rather
high. A reliable estimation of the block entropies for $n>30$ is still
an open question. The results for the entropy of the two books suggest
in agreement with Shannon`s data and Hilberg`s findings that the mean
entropy per letter decays to its limit according to a square root law.
As a consequence the number of different subwords in texts increases
with the number of letters $n$ according to a stretched exponential
law.  Our estimations for the growth yield for $n=100$ a total number
of about $2^{53}$ different subwords.  Most of the subwords which
would be possible from the combinatorial point of view are actually
forbidden and do not appear in real texts.  We investigated also how
the number of genuine English words $N(L)$ (formally defined here as
sequences of letters between spaces and/or marks) increases with the
length $L$ of a text. For Grimm's Tales we found the scaling law
$N(L)=22.8 \cdot L^{0.46}$, i.e. reading the book we find permanently
new words.
\par
More empirical data on long texts and further studies of the statistical
effects due to finite length of the samples are needed in order to reach 
a more definite conclusion about the scaling properties.

\stars
The authors thank K.~Albrecht, T.~Boseniuk, J.~Freund, H.~Herzel,
G.~Nicolis and A.~Schmitt for fruitful discussions. Further we thank
the {\em Project Gutenberg Etext} at Illinois Benedictine College for
providing the ASCII-files of the investigated literature.


\end{document}